\documentclass[12pt]{article}

\usepackage{graphicx}  
\usepackage[utf8]{inputenc}
\usepackage{authblk}
\usepackage{abstract}
\usepackage{setspace}
\usepackage{bm}        
\usepackage{amssymb}  
\usepackage{amsmath}
\usepackage{indentfirst}
\usepackage[font=footnotesize,labelfont=bf]{caption}
\usepackage{booktabs}

\usepackage[margin=1 in]{geometry} 

\usepackage[
backend=biber,
style=nature,
sorting=none
]{biblatex}
\title{\vspace{-12mm}\large \textbf{Nonlinear Thomson scattering with ponderomotive control}}

\author[1]{D. Ramsey\thanks{dram@lle.rochester.edu}$^,$}     
\author[2]{B. Malaca}
\author[3]{A. Di Piazza}
\author[3]{M. Formanek}
\author[1]{P. Franke}
\author[1]{D.H. Froula}
\author[2]{M. Pardal}
\author[1]{T.T. Simpson}
\author[2]{J. Vieira}
\author[1]{K. Weichman}
\author[1]{J.P. Palastro\thanks{jpal@lle.rochester.edu}$^,$} 
\affil[1]{
University of Rochester, Laboratory for Laser Energetics, Rochester, New York, 14623 USA
}
\affil[2]{
GoLP/Instituto de Plasmas e Fusão Nuclear, Instituto Superior Técnico, Universidade de Lisboa, Lisbon, Portugal
}
\affil[3]{
Max-Planck-Institut für Kernphysik, Saupfercheckweg 1, D-69117 Heidelberg, Germany
}

\newcommand{\gamPsqAvg}{\ensuremath{\langle \gamma_\perp^2 \rangle} }        

\newcommand{\PZavg}{\ensuremath{\langle u_z \rangle} }                       

\newcommand{\gamAvg}{\ensuremath{\langle \gamma \rangle} }                   

\newcommand{\appropto}{\mathrel{\vcenter{                                    
  \offinterlineskip\halign{\hfil$##$\cr
    \propto\cr\noalign{\kern2pt}\sim\cr\noalign{\kern-2pt}}}}}
\date{}             

\begin{document}
\maketitle
\doublespacing      
\vspace{-12mm}
\begin{center}
    \textbf{Abstract}
\end{center}
\vspace{-3mm}

In nonlinear Thomson scattering, a relativistic electron reflects and re-radiates the photons of a laser pulse, converting optical light to x rays or beyond. While this extreme frequency conversion offers a promising source for probing high-energy-density materials and driving uncharted regimes of nonlinear quantum electrodynamics, conventional nonlinear Thomson scattering has inherent tradeoffs in its scaling with laser intensity. Here we discover that the ponderomotive control afforded by spatiotemporal pulse shaping enables novel regimes of nonlinear Thomson scattering that substantially enhance the scaling of the radiated power, emission angle, and frequency with laser intensity. By appropriately setting the velocity of the intensity peak, a spatiotemporally shaped pulse can increase the power radiated by orders of magnitude. The enhanced scaling with laser intensity allows for operation at significantly lower electron energies and can eliminate the need for a high-energy electron accelerator.


\newpage
\section*{\fontsize{14}{16.8}\selectfont Introduction}
\vspace{-2mm}

Bright sources of high-energy photons lead to advancements in a range of disciplines, including ultrafast biology and material science, nonlinear quantum electrodynamics, nuclear spectroscopy, and radiotherapy \cite{Schoenlein2237,Pfeiffer2006,Gaffney1444,Barends445,Barty2008,Clark56,RINGWALD2001107,AlkoferPairs,HastingsMossb,SetoNuc,Suortti_2003,Biston2317,MONTAYGRUEL2018582}. The brightest sources currently reside at large accelerator facilities in the form of x-ray free-electron lasers or synchrotrons \cite{Emma2010,Tavella2011,Castelvecchi2015}. While laser-driven sources \cite{sarachick1970classical,SprangeSynch,esarey1993nonlinear,chen1998experimental,rousse2004production,Seres2005,SprangleFEL,Fuchs2009,Cipiccia2011,TaPhuoc2012,CordeREMP,Albert_2014,Powers2014,Andriyash2014,Rykovanov,sarri2014ultrahigh,PalastroBeta,ClarkMeV,Yan2017,Vieira2021} promise a smaller-scale, widely accessible alternative, challenges in achieving the required photon number, energy, and coherence have held these sources back. Of the potential candidates, nonlinear Thomson scattering (NLTS) can produce extremely high energy, collimated radiation in a relatively controlled setting \cite{SprangeSynch,esarey1993nonlinear, chen1998experimental,TaPhuoc2012,Powers2014,sarri2014ultrahigh,Yan2017}. Like the other candidates, however, NLTS has inherent constraints that currently impede its realization as a practical light source. 

\begin{figure}
\centering\includegraphics[width=3.5in]{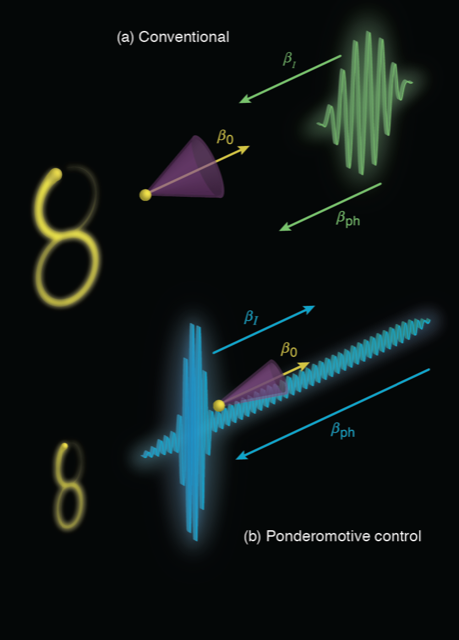}
\caption{(a) A conventional NLTS configuration in which the intensity peak and phase fronts of a laser pulse travel in the opposite direction of the electron. At the rising edge of the intensity peak, the ponderomotive force decelerates the electron, redshifting the emitted frequencies and widening their emission angle (purple cone). (b) NLTS with ponderomotive control aligns the velocities of the intensity peak and the electron. Here the ponderomotive force of the intensity peak increases or maintains the electron velocity, allowing for higher-frequency emission into a smaller angle. The electron trajectory in its average rest frame (figure-eight motion) is depicted to the left of each case.}
\label{fig:f1}
\end{figure}

In NLTS, a relativistic electron collides with a laser pulse traveling in the opposite direction [Fig. \ref{fig:f1}(a)]. The electron rapidly oscillates in the fields of the pulse, reflecting and re-radiating the incident photons. The properties of the radiation depend on the vector potential ($a$) and frequency ($\omega_0$) of the pulse and the initial electron energy ($\gamma_0$) (energy and charge are normalized to $m_ec^2$ and $e$ throughout). Maximizing the radiated power ($P$), or the number of photons, requires large vector potentials ($P \propto a^2$). In these strong fields ($a \gg 1 $), the ponderomotive force of the pulse appreciably decelerates the electron and increases the amplitude of its oscillations along the direction of its initial motion \cite{esarey1993nonlinear, sarachick1970classical}. This redshifts the emitted frequencies ($\omega_n$) and widens the emission angle ($\theta_e$): $\omega_n \approx 8n\gamma_0^2\omega_0/a^2 $ and $\theta_e \sim a/\gamma_0$, where $n$ is an integer \cite{esarey1993nonlinear}. This tradeoff between the power, spectrum, and emission angle constrains the utility of NLTS. 

Spatiotemporal pulse shaping provides control over the ponderomotive force, which can compensate the ponderomotive deceleration in NLTS \cite{kondakci2017diffraction,sainte2017controlling,froula2018spatiotemporal,palastro2020dephasingless,simpson2020nonlinear}. As an example, the chromatic aberration of a diffractive optic and a chirp can be used to control the location and time at which each temporal slice within a pulse comes to its focus, respectively. By adjusting the chirp, the resulting intensity peak, and therefore the ponderomotive force, can travel at any velocity, either forward or backward with respect to the phase fronts, over distances much longer than a Rayleigh range \cite{sainte2017controlling,froula2018spatiotemporal}. Aside from extending the interaction length, a ponderomotive force that counter-propagates with respect to the phase fronts can \textit{accelerate} an electron in NLTS \cite{ramsey2020vacuum} and provide unique insight into the corresponding quantum process, i.e., nonlinear Compton scattering \cite{di2021scattering}. 

Here we describe novel regimes of nonlinear Thomson scattering that exploit the ponderomotive control afforded by spatiotemporal pulse shaping to substantially enhance the scaling of power, emission angle, and frequency with laser intensity. For high-intensity pulses ($a^2 \gg 1 $), these regimes exhibit orders-of-magnitude higher radiated powers and smaller emission angles than conventional NLTS. Further, the improved scaling with laser intensity allows for lower electron energies, relaxing the requirements on the electron accelerator. While the results can be generalized to any ponderomotive velocity ($\beta_I = v_I/c$), we focus on two regimes: ``drift-free" NLTS, which preserves spectral properties independent of the laser pulse shape and intensity, and ``matched" NLTS, which offers a spectrum that can be tuned independently of the initial electron energy.

\section*{\fontsize{14}{16.8}\selectfont Results}
\vspace{-2mm}

Figure \ref{fig:f1} contrasts backscattering configurations for conventional NLTS and NLTS with ponderomotive control (NLTSPC). Conventional NLTS employs a standard laser pulse with an intensity peak and phase that counter-propagate at the vacuum speed of light with respect to a relativistic electron. NLTSPC employs a spatiotemporally shaped pulse with an intensity peak that counter-propagates with respect to its phase fronts and co-propagates with respect to the electron. In both cases, as the electron enters the leading edge of the intensity peak, it begins oscillating in the polarization (transverse) and propagation (longitudinal) directions. For a linearly polarized pulse, this motion, in a frame moving with the average longitudinal velocity of the electron, traces out the characteristic figure eight of NLTS. While both cases exhibit this qualitative motion, NLTSPC provides additional freedom over the electron trajectory. 

The electron trajectory evolves in response to the vector potential $\textbf{a} =a(z - \beta_I t)\cos(z+t)\hat{\textbf{x}}$, where the envelope $a$ captures the motion of the intensity peak and time and space have been normalized to $\omega_0$ and $\omega_0/c$, respectively. With the recognition that the vector potential changes slowly with the coordinate $\xi = z-\beta_It$ and rapidly with $\eta = z+t$, a multiple time-scale analysis (see \textbf{Methods}) reveals the local conservation equation $\partial_{\eta}(\gamma+u_z) = 0$ or $\gamma+u_z = h(\xi)$, where $\textbf{u}$ is the electron momentum. This relation indicates that the Hamiltonian ($h$) of the electron in a frame moving with the phase velocity  depends only on the slow coordinate $\xi$. Using this relation, one can show that
\begin{equation}
    h(\xi) = \gamAvg(1-\beta_I^{-1})+\gamma_0(\beta_0-\beta_I^{-1}),
\end{equation}

\noindent where $\gamAvg = \gamma_I^2 \gamma_0 (1-\beta_I \beta_0) - \beta_I \gamma_I^2[\gamma_0^2(1-\beta_I \beta_0)^2-\gamma_I^{-2}\gamPsqAvg]^{1/2} $ is the electron energy averaged over a cycle of the laser pulse, $\beta_0 = (1-\gamma_0^{-2})^{1/2}$ is the initial longitudinal velocity of the electron, $\gamma_I^2 = (1-\beta_I^2)^{-1}$, and $\gamPsqAvg = 1+\frac{1}{2}a^2$. Note that $h$ depends only on the initial electron energy, the ponderomotive velocity, and the local value of the vector potential. From here on, the $\xi$-dependence of all quantities that depend on $a$ is understood.

The Hamiltonian ($h$) determines all details of the electron trajectory and the radiation properties. Specifically, the radiation results from the time-dependent curvature of the electron trajectory, which is set by the amplitude of the transverse ($x_0$) and longitudinal ($z_0$) oscillations, $x_0 = a/h$ and $z_0 = a^2/8h^2$, about a drift motion characterized by the longitudinal velocity, $\beta_d = (h^2-\gamPsqAvg)/(h^2+\gamPsqAvg)$  \cite{esarey1993nonlinear}. The cycle-averaged power $(\langle P \rangle)$, emission angle with respect to the initial electron velocity, frequency of each harmonic, and bandwidth ($\omega_b$) all depend on $h$: $\langle P \rangle = r_eh^2 a^2$/3, $\theta_e  \backsim  a/h$, $\omega_n = nh^2/\gamPsqAvg$, and $\omega_b = 3a^3h^2/4\gamPsqAvg$, where $r_e$ is the classical electron radius. Conventional NLTS corresponds to the special case of $\beta_I = -1$ and $h = (1+\beta_0)\gamma_0$ with the radiation properties found in Table \ref{tab:RadiationProperties}. Through $h$, the ponderomotive velocity ($\beta_I$) provides an additional parameter to tune the trajectory and radiation properties in NLTSPC.

\begin{table}
\setlength{\tabcolsep}{16pt}
\centering
 \begin{tabular}{c c c c}

    \toprule 
                               &  Conventional                                  & Drift-free                                                        & Matched  \\ \midrule
                               
    $\langle P \rangle$        & $\frac{r_e}{3}(1+\beta_0)^2\gamma_0^2 a^2$     & $\frac{r_e}{3}(1+\beta_0)^2\gamma_0^2(1+\tfrac{1}{2}a^2)a^2$              & $\frac{r_e}{3}(1+\beta_I)^2\gamma_I^2 (1+\tfrac{1}{2}a_0^2) a_0^2$\\
    
    \\
     
    $\theta_e$                 & $\dfrac{a}{(1+\beta_0)\gamma_0}$               & $\dfrac{a}{(1+\beta_0)\gamma_0(1+\tfrac{1}{2}a^2)^{1/2}}$          & $\dfrac{a_0}{(1+\beta_I)\gamma_I(1+\tfrac{1}{2}a_0^2)^{1/2}}$\\   
   
    \\
    
    $\omega_n$                 & $\dfrac{n(1+\beta_0)^2\gamma_0^2}{1+\tfrac{1}{2}a^2}$        & $n(1+\beta_0)^2\gamma_0^2$                             & $n(1+\beta_I)^2\gamma_I^2$ \\
   
    \\
    
    $\omega_b$                 & $\dfrac{3(1+\beta_0)^2\gamma_0^2a^3}{4(1+\tfrac{1}{2}a^2)}$    & $\frac{3}{4}a^3 (1+\beta_0)^2\gamma_0^2$             & $\frac{3}{4}a_0^3 (1+\beta_I)^2\gamma_I^2$ \\             
           
    \bottomrule

\end{tabular}
\caption{\label{tab:RadiationProperties} The radiation properties: cycle-averaged power ($\langle P \rangle$), emission angle ($\theta_e$), harmonic frequency ($\omega_n$), and bandwidth ($\omega_b$) for conventional, drift-free, and matched NLTS. For matched NLTS, it has been assumed that the electron spends most of the interaction at an $a \approx a_0$.}
\end{table}

\begin{figure}
\centering\includegraphics[width=3.5in]{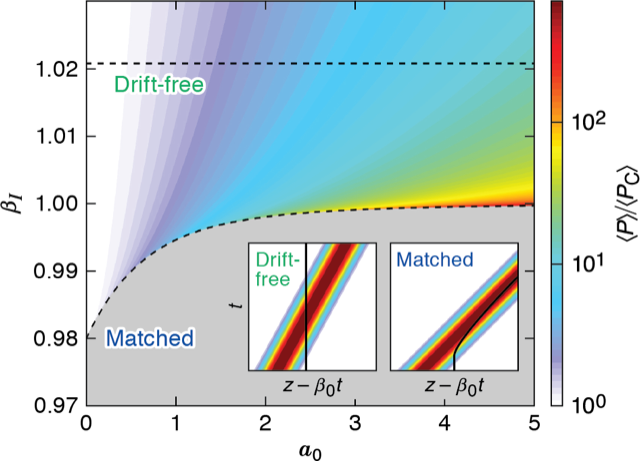}
\caption{The cycle-averaged radiated power as a function of the ponderomotive velocity ($\beta_I$) and the vector potential ($a$) normalized to power radiated in conventional NLTS ($ \langle P_C \rangle $). Here $\gamma_0 = 5$, and, for the purpose of calculating $\langle P_C \rangle$, $\beta_I = -1$ . The dashed lines indicate the ``matched" and ``drift-free" conditions of NLTSPC. Within the gray region, the ponderomotive force of the intensity peak accelerates the electron to a velocity greater than $\beta_I$, and the electron outruns the intensity peak---a situation not considered here \cite{ramsey2020vacuum}. The insets depict the cycle-averaged electron trajectories (black lines) relative to the motion of the intensity peak (contours) for the drift-free and matched regimes.}
\label{fig:f2}
\end{figure}

Figure \ref{fig:f2} illustrates the impact of using the ponderomotive velocity to tune the electron trajectory. For ponderomotive velocities parallel to the initial electron velocity, the radiated power can be orders-of-magnitude larger than in conventional NLTS ($\langle P_C \rangle$). Further, the enhancement in the radiated power increases with the vector potential ($a$), favoring high-intensity laser pulses. Note that here and throughout, the parameters have been chosen to ensure that NLTS occurs in the classical regime, i.e., $\hbar\omega_b \ll \gamma_0$ \cite{di2012systems}. 

To understand why the properties of the radiation change with the ponderomotive velocity ($\beta_I$), consider the longitudinal drift velocity of the electron  ($\beta_d$). The velocity can increase or decrease as the electron enters the intensity peak, depending on the value of $\beta_I$. In conventional NLTS, $h = (1+\beta_0)\gamma_0$ is independent of $a$. As a result, the increase in $\gamPsqAvg$ as the electron enters the pulse necessarily decreases $\beta_d$. Said differently, the electron is ponderomotively decelerated by the counter-travelling intensity peak of the pulse. In NLTSPC, $\beta_I \approx 1$ and $h(\xi)$ depends on $a$. Now $h(\xi)$ and, as a result, $\beta_d$ increase as the electron enters the intensity peak, i.e., the electron is ponderomotively accelerated by the co-travelling intensity peak of the pulse. In fact, this acceleration can become so large that the electron outruns the intensity peak altogether (gray area in Fig. \ref{fig:f2}). An increase in $\beta_d$, and therefore $h$, enhances the scaling of the radiation properties ($\langle P \rangle, \omega_n, \omega_b \propto h^2$ and $\theta_e \propto h^{-1}$). Two specific cases illustrate this benefit more clearly.

\begin{figure} 
\centering\includegraphics[width=3.5in]{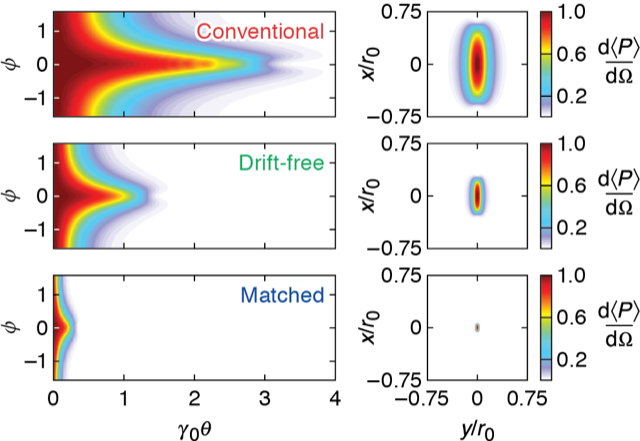}
\caption{(left) Power radiated per steradian as a function of the angle with respect to the initial electron velocity ($\theta$) and the angle coplanar with the laser polarization ($\phi$) for a $\gamma_0 = 5$ and a maximum vector potential $a_0 = 3$. (right) the projection of the radiated power on a plane located an arbitrary distance $r_0$ from the source. For clarity, each has been normalized to its maximum value: 0.016, 0.50, 204  MeV$\cdot$sr\textsuperscript{-1}$\cdot$ps\textsuperscript{-1} for conventional, drift-free, and matched NLTS, respectively.}
\label{fig:f3}
\end{figure}

\begin{figure}
\centering\includegraphics[width=3.5in]{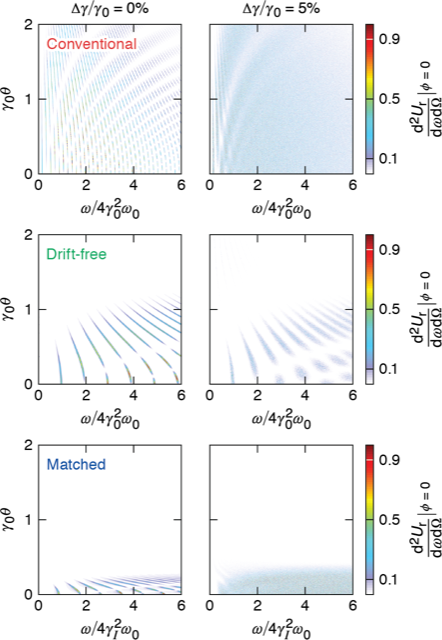}
\caption{ Spectrum of emitted radiation from a collection of electrons with $\gamma_0 = 5$ oscillating in a laser pulse with a maximum vector potential $a_0 = 3$. On the left, the electrons have no energy spread, while on the right, the electrons have a $\Delta \gamma/\gamma_0 =5 \%$ spread all in the longitudinal momentum. The quantity $U_r$ has units of energy and each plot is normalized to its maximum value. Note that for matched NLTS the horionzontal axis is scaled to $4\gamma_I^2\omega_0$, not $4\gamma_0^2\omega_0$, and therefore extends to much higher, absolute frequencies.}
\label{fig:f4}
\end{figure}

``Drift-free" NLTS employs a superluminal intensity peak to compensate the ponderomotive deceleration of the electron. When $\beta_I=\beta_0^{-1}$, the ponderomotive force of the intensity peak increases the energy and the longitudinal momentum of the electron in the right balance to maintain a constant longitudinal drift velocity ($\beta_d$) throughout the interaction (Fig. \ref{fig:f2} inset). The resulting value of $h=(1+\beta_0)\gamma_0\gamPsqAvg^{1/2}$ provides the radiation properties displayed in Table \ref{tab:RadiationProperties}. Each property has an improved scaling with laser intensity ($a^2$) when compared to conventional NLTS. Aside from the enhanced power (Fig. \ref{fig:f2}), the radiation in drift-free NLTS is emitted into a much smaller angle when $a\gg1$ (c.f., Fig. \ref{fig:f3}). Further, regardless of the time-dependent vector potential experienced by the electron, the harmonic frequencies remain fixed (Table \ref{tab:RadiationProperties}). Compensating the ponderomotive deceleration of the electron eliminates the redshift ($\gamPsqAvg$ factor) in the emitted harmonics ($\omega_{n}$) and bandwidth ($\omega_{b}$), mitigating the tradeoff between the power, emitted angle, and spectrum (Fig. \ref{fig:f4}). 

``Matched" NLTS (subscript M) uses a subluminal intensity peak to ponderomotively accelerate the electron. Here the intensity peak intercepts the electron from behind and gradually accelerates it to an asymptotic velocity of $\beta_d = \beta_I$. This allows the electron to experience a near-constant vector potential for an extended distance (Fig. \ref{fig:f2} inset). Setting the ponderomotive velocity to satisfy $(\beta_I -\beta_0)\gamma_I\gamma_0 = a_0/\sqrt{2}$ ensures that the electron co-travels with the intensity peak near the maximum vector potential ($a_0$). With this condition met, $h=(1+\beta_I)\gamma_I\gamPsqAvg^{1/2}$ yielding the radiation properties found in Table \ref{tab:RadiationProperties}.

Matched NLTS represents the optimal case of NLTSPC. With a smaller ponderomotive velocity, the accelerated electron would outrun the intensity peak; with a larger, but still subluminal, ponderomotive velocity, the pulse would eventually overtake and outrun the accelerated electron, limiting the interaction length \cite{ramsey2020vacuum,mendoncca2007reflection}. For large vector potentials ($a_0\gg1$), the optimal scalings (Table \ref{tab:RadiationProperties}) result in a radiated power far greater and an emission cone far narrower than either drift-free or conventional NLTS (Figs. \ref{fig:f2} and \ref{fig:f3}). In this limit, $\gamma_I \simeq \sqrt{2}a_0 \gamma_0$, such that $\langle P \rangle \propto a_0^6$, $\theta_e \propto a_0^{-1}$, $\omega_{n} \propto a_0^2$, and $\omega_{b} \propto a_0^5$. 

While NLTS typically involves the intersection of a pre-accelerated, highly relativistic electron ($\gamma_0 \gg 1$) with an intense laser pulse ($a^2 \gg 1$), matched NLTS allows for a different paradigm: a spatiotemporally shaped laser pulse can both accelerate and scatter from an initially non-relativistic electron ($\gamma_0 \gtrsim 1$). The condition on the ponderomotive velocity can be adjusted to accommodate small electron energies, i.e., $\gamma_I \approx \sqrt{2}a_0$. By substituting a preliminary accelerator for a high-intensity laser pulse, this unique regime of NLTS would trade the favorable scaling of the radiation properties with $\gamma_0$ for the enhanced scaling with $a_0$. For instance, an initially non-relativistic electron accelerated to a drift velocity $\beta_d = \beta_I$ by an intensity peak with $a_0=6.3$ could radiate the same power as an electron with $\gamma_0 = 500$ in a standard pulse with $a_0 = 0.5$.

In both conventional NLTS and NLTSPC, a single electron radiates a coherent spectrum with well-defined peaks (Fig. \ref{fig:f4}). A practical light source requires a large number of photons, which can only be achieved by using an electron bunch. Depending on the source, these bunches can have non-ideal features, such as spatial and momentum spreads. Simulations of NLTS from electron bunches using the particle-in-cell code OSIRIS \cite{FonsecaOSIRIS} with the newly developed package RaDiO \cite{Vieira2021} (see \textbf{Methods}) showed that, in a plane-wave pulse, the spatial width of the electron bunch had no discernible effect on the spectrum. Similarly, a longitudinal momentum spread equivalent to a $\Delta \gamma/\gamma_0 = 1\%$ had little effect. A momentum spread of $\Delta \gamma/\gamma_0 = 5\%$ was, however, sufficient to blur each spectral peak (Fig. \ref{fig:f4}). For conventional  NLTS, the proximity of the harmonics combined with the blurring ($\Delta \omega_n \sim 2\omega_n\Delta\gamma/\gamma_0)$ created a near-continuous spectrum formed by the overlap of adjacent harmonic peaks. For drift-free NLTS, the relatively large separation between the harmonics mitigated this overlap. Matched NLTS also produces harmonics with a relatively large separation. However, electrons with momentum that do not satisfy the matching condition can experience vastly different vector potentials along their trajectory. This contributes an additional blurring to each harmonic. As a result, the matched NLTS spectrum also exhibited a near-continuous spectrum. 

\section*{\fontsize{14}{16.8}\selectfont Discussion}
\vspace{-2mm}

Nonlinear Thomson scattering with ponderomotive control can produce extremely high energy photons with a spectrum that can be tuned through the initial electron energy, the laser amplitude, and now, the ponderomotive velocity. The added flexibility enabled by ponderomotive control eliminates the tradeoffs inherent to conventional NLTS. Specifically, the intensity peak of a laser pulse in conventional NLTS counter-propagates with respect to the electron, causing a ponderomotive deceleration that redshifts the radiated frequencies and widens their angle of emission---an effect exacerbated by large laser intensities. Instead, by reversing the propagation direction of the intensity peak with respect to the phase fronts, a spatiotemporally shaped pulse can ponderomotively accelerate the electron. For drift-free NLTS with $a_0\gg1$, this removes the intensity dependence of the harmonic frequencies and emission angle, while enhancing the intensity scaling of the bandwidth and power. For matched NLTS with $a_0\gg1$, this enhances the intensity scaling of all the radiation properties.

The enhanced scalings with laser intensity switch the burden in NLTS from the accelerator to the laser---a situation ideal for existing and emerging high-energy, high-power laser facilities. Consider a $\lambda_0 = 1 \, \mathrm{\mu m}$ wavelength laser pulse ($\lambda_0 = 2\pi c/\omega_0$): Drift-free NLTS with a laser intensity $I = 3.4\times10^{19}$ W/cm\textsuperscript{2} ($a_0 = 5$) and a 10 MeV ($\gamma_0 = 20$) electron bunch produces the same radiated power and an emission angle  comparable to conventional NLTS with a laser intensity of $I = 1.4\times10^{18}$ W/cm\textsuperscript{2} ($a_0 = 1$) and a 180 MeV ($\gamma_0 = 350$) electron bunch. For these parameters, drift-free has a smaller harmonic frequency but comparable overall bandwidth. At an identical laser intensity ($a_0=5$), matched NLTS can radiate the same power into a narrower emission angle and broader bandwidth with only a 1.5 MeV electron bunch. 

Co-propagation of the electron and intensity peak in NLTSPC extends the maximum interaction length ($L$) and duration of the radiation [$t_{\mathrm{r}}\sim(1-\beta_d)L$] beyond that of conventional NLTS, which increases the total radiated energy ($U_{\mathrm{r}} \sim LP$). For an intensity peak of duration $\tau$, $L_{C} \sim \tau/2$ and $L_{D} \sim  \gamma_0^2\tau$ in conventional and drift-free NLTS, respectively. Matched NLTS requires an initial distance $L_{M} \sim a_0^2\gamma_0^2\tau$ for the intensity peak to accelerate the electron to its approximate asymptotic velocity. In principle, once this velocity is reached, the interaction length is unbounded. The extended interaction lengths in NLTSPC do, however, come with a caveat: a spatiotemporally shaped pulse must have a larger energy ($U_{\mathrm{pulse}} \propto L$) to sustain its intensity over the maximum interaction distance. Of course, shorter interaction lengths can always be used. 

NLTS has the potential to provide a compact, low-cost alternative to light sources based on conventional accelerators with far-reaching benefits in medicine and basic science. For fundamental physics, in particular, NLTS represents one of the optimal configurations for accessing the strong-field regime of quantum electrodynamics (QED). The photons emitted during NLTS can scatter from multiple laser photons and decay into electron--positron pairs (i.e., undergo Breit--Wheeler pair production). These decays can trigger a cascade of photon, positron, and electron creation, producing conditions relevant to extreme astrophysical objects or unique states of matter characterized by the interplay of collective plasma and high-field quantum processes. Proximity to the nonlinear regime of Breit--Wheeler is quantified by the invariant parameter $\chi = |F^{\mu\nu}k_{\nu}|/E_{cr}$, where $F^{\mu \nu}$ is the electromagnetic tensor, $k$ the photon four-momentum, $E_{cr} = m_ec^2/\hbar\omega_0$, and $\chi \approx 1$ roughly marks the transition from the linear to nonlinear regimes \cite{di2012systems}. The increase of the emitted frequencies in the new NLTS schemes described here enhance the scaling of $\chi$ with respect to $a_0$, providing access to the strong-field QED regime at either smaller values of $a_0$ or $\gamma_0$. Setting the photon energy to $\hbar \omega_b$, the expressions for $\chi$ in the limit $a_0\gg1$ are given by $\chi_C = 12 E_{cr}^{-2} \gamma_0^2a_0^2$, $\chi_D = 6 E_{cr}^{-2} \gamma_0^2a_0^4$, and $\chi_M = 12 E_{cr}^{-2} \gamma_0^2a_0^6$. As an example, for a $\lambda_0 = 1 \, \mu$m wavelength laser pulse with $a_0 = 100$ and $\gamma_0 = 20$, $\chi_C = 3\times10^{-4} $ and $\chi_D = 1.4$. With ponderomotive control, even a mildly relativistic electron beam could place photon--photon scattering in the strong-field regime of QED.

\section*{\fontsize{14}{16.8}\selectfont Methods}
\vspace{-2mm}

The radiation emitted in nonlinear Thomson scattering (NLTS) results from the acceleration of an electron in the fields of an intense laser pulse. These fields were modeled using a vector potential ($\textbf{A}$) that captures the salient features of a spatiotemporally shaped pulse: $\textbf{a} =a(z - \beta_I t)\cos(z+t)\hat{\textbf{x}}$, where $\textbf{a} = e \textbf{A}/m_e c$, $a(z-\beta_It)$ represents an envelope traveling at the ponderomotive velocity $\beta_I$, and time and space have been normalized to $\omega_0$ and $\omega_0/c$, respectively. In response to the fields of the laser pulse, the electron momentum ($\textbf{u}$) and energy ($\gamma$) evolve according to the equations of motion:
\begin{subequations} 
\begin{align}
         \frac{d\textbf{u}}{dt} &= \frac{\partial \textbf{a}}{\partial t} - \frac{\textbf{u}}{\gamma} \times ( \nabla \times \textbf{a} ), \tag{M1} \\
        \frac{d\gamma}{dt} &= -\frac{\textbf{u}}{\gamma} \cdot \frac{\partial \textbf{a}}{\partial t}, \tag{M2}
\end{align}
\end{subequations}
\noindent where momentum and energy have been normalized to $m_ec$ and $m_ec^2$, respectively. For an electron initially outside of the pulse envelope with no transverse momentum, the transverse component of Eq. (M1) provides $u_x = \text{a}$, where $\text{a} = \hat{\textbf{x}}\cdot\textbf{a}$. 

The coordinate transformations $\eta = z + t$ and $\xi = z -\beta_I t$ facilitate analysis of the longitudinal momentum and energy. In terms of these coordinates, $\textbf{a} =a(\xi)\cos(\eta)\hat{\textbf{x}}$, and 
\begin{subequations} \label{coordSwitch}
    \begin{align}
         \left[ \frac{\partial}{\partial \eta} +\left( \frac{\beta_z-\beta_I}{1+\beta_z} \right) \frac{\partial}{\partial \xi} \right]u_z &= -\frac{1}{2(1+\beta_z)\gamma}\left(\frac{\partial}{\partial \eta} + \frac{\partial}{\partial \xi} \right)\text{a}^2, \label{coordSwitchP} \tag{M3} \\
         \left[  \frac{\partial}{\partial \eta} +\left( \frac{\beta_z-\beta_I}{1+\beta_z} \right) \frac{\partial}{\partial \xi} \right]\gamma &= -\frac{1}{2(1+\beta_z)\gamma}\left(\frac{\partial}{\partial \eta} -\beta_I \frac{\partial}{\partial \xi} \right)\text{a}^2, \label{coordSwitchGAM} \tag{M4}
    \end{align}
\end{subequations}

\noindent where $\beta_z = u_z/\gamma$. For a typical spatiotemporally shaped pulse, the duration of the intensity peak is much longer than the optical period, i.e., $|\partial_\eta \textbf{a}| \gg |\partial_\xi \textbf{a}|$. Further, the electron travels at a relativistic velocity ($\beta_z \lesssim 1$) that is antiparallel to the phase velocity of the pulse. Together, these allow for an approximate solution to Eqs. \eqref{coordSwitchP} and \eqref{coordSwitchGAM} based on a multiple-time-scale approach, i.e., using $|\partial_\eta| \gg |\partial_\xi|$.

To lowest order, one finds the local conservation equation $\partial_\eta (\gamma + u_z) = 0$. Integrating this equation provides the local Hamiltonian ($h$) of the electron in a frame moving with the phase velocity, $\gamma + u_z = h(\xi)$, or, equivalently, $\gamAvg + \PZavg= h(\xi)$, where $\langle \rangle$ denotes an average over the rapidly varying phase of the laser pulse. To next order, one finds the slowly varying conservation equation $\partial_\xi (\gamma - \beta_I u_z) = 0$, which, upon phase averaging, becomes $\partial_\xi (\gamAvg - \beta_I \PZavg) = 0$ \cite{ramsey2020vacuum,mckinstrie1996electron}. For an electron with an initial momentum $\mathrm{\bold{u}_0}=|\beta_0|\gamma_0 \hat{\textbf{z}}$, $\gamAvg - \beta_I \PZavg = \gamma_0(1 -\beta_I \beta_0)$. Using this to eliminate $\PZavg$ in $h$ yields Eq. (1).

With an expression for $h$, one can follow the derivation presented by Esarey \textit{et al.} \cite{esarey1993nonlinear} for the electron trajectory and radiation properties; however, unlike Esarey \textit{et al.}, these properties are dynamic, varying with $\xi$ through the dependence of $h$ on $a(\xi)$. For brevity in notation, reference to the explicit $\xi$ dependence of slowly varying quantities will now be dropped. From $\gamma + u_z = h$ and $u_x = \text{a}$, one can find the transverse and longitudinal velocities: $\beta_x = \text{a}/\gamma$ and $\beta_z = (h^2 -1-\text{a}^2)/(h^2 + 1+\text{a}^2)$. The electron coordinates then evolve according to $\dot{x} = \text{a}/h$ and $\dot{z} = \frac{1}{2}(1-h^{-2}-\text{a}^2)$, where $\cdot$ represents differentiation with respect to $\eta$ and $d \eta = (1+\beta_z)dt$ has been used. Upon integrating,
\begin{subequations} \label{Orbits}
    \begin{align}
       x(\eta) &=  x_0 \sin(\eta), \label{transTraj} \tag{M5} \\
       z(\eta) &=  \frac{1}{2}\left(1- \frac{\gamPsqAvg}{h^2}\right)\eta - z_0 \sin(2\eta), \label{axialTraj} \tag{M6}
    \end{align}
\end{subequations}
to lowest order in the time-scale expansion \noindent, where $x_0= a/h$, $z_0 = a^2/8h^2$, and $\gamPsqAvg = 1+\frac{1}{2}a^2$. The oscillating terms trace out a figure eight---a general feature of electron motion in NLTS. The amplitude of these oscillations ($x_0$, $z_0$) depends on the local values of $a$ and $h$. Substituting $\eta = z+t$ into Eq. \eqref{axialTraj}, rearranging terms, and differentiating with respect to $t$ provides the longitudinal drift velocity, $\beta_{d} = (h^2-\gamPsqAvg)/(h^2+\gamPsqAvg)$.

The local values of $a$ and $h$ determine all properties of the emitted radiation. Acting like a nonlinear, relativistic moving mirror, the electron emits harmonics of the twice-Doppler upshifted frequency of the laser pulse:
\begin{equation} \label{Gen_DS}
    \omega_n = n\frac{1+\beta_d}{1-\beta_d} = n\frac{h^2}{\gamPsqAvg}. \tag{M7}
\end{equation} 
\noindent The harmonics range over a bandwidth characterized by the invariant critical integer $n_c = \frac{3}{4}a^3$, such that $\omega_b = \frac{3}{4}a^3\omega_1$. The radiation is spread over an angle ($\theta_e$) determined by the bounds of the oscillations: $ \theta_e  \backsim  |z_0/ x_0|  =\frac{1}{8}ah^{-1}$. Finally, the phase-averaged radiated power ($ \langle P \rangle$) is calculated using the relativistic Larmor formula, which yields $ \langle P \rangle    = \frac{r_e}{3} h^2a^2$,  where $r_e$ is the classical electron radius. For an alternative derivation of the radiation properties, one can compute the curvature of the electron trajectory from Eqs. (M5) and (M6) and use the general expressions found in Ref. \cite{jackson2007electrodynamics}.  

For Figs. \ref{fig:f2} and \ref{fig:f3}, the radiation properties were calculated using the peak vector potential ($a_0$). To calculate the angular distribution of the radiated power (Fig. \ref{fig:f3}), Eq. (36) in Ref. \cite{esarey1993nonlinear} was integrated over frequency. The resulting expression, a summation over an infinite number of harmonics, is proportional to the duration of the interaction. This duration was divided out, providing an expression for the power instead of the energy. The summation was performed numerically by truncating at $n_{\mathrm{max}} = 4n_c = 3 a_0^3$. Additional summations performed with larger values of $n_{\mathrm{max}}$ confirmed that harmonics beyond $4n_c$ have negligible effect on the result. When integrated over solid angle ($\Omega$), the resulting power was in excellent agreement with independent calculations that used the relativistic Larmor formula. Further, the angular distribution was compared to calculations that used electrons trajectories directly in the Liénard–Wiechert potential. The electron motion was evolved in the model vector potential using the algorithm detailed in Ref. \cite{gordon2021special}, while the Liénard–Wiechert potential was calculated using the algorithm described in Ref. \cite{thomas2010algorithm}. At different angles, the resulting spectrum was integrated over frequency and divided by the duration of the interaction confirming the distributions displayed in Fig. \ref{fig:f3}.

The 2D OSIRIS \cite{FonsecaOSIRIS} simulations were performed in a moving frame using a non-evolving laser pulse. Specifically, the transverse vector potential was given by $\textbf{a} =a(z - \beta_I t)\cos(z+t)\hat{\textbf{x}}$. The shape function ($a$) had a constant flattop region where $a=a_0$ surrounded by symmetric rising and falling edges defined by a smooth fifth order ramping polynomial. For simulations of NLTSPC, the rise and fall times of the pulse were $5$ and the total length was $40$ (length and time are in units of $c\omega_0^{-1}$ or $\omega_0^{-1}$). For conventional NLTS, a longer pulse was used to ensure a comparable interaction length; the rise and fall times were $100$ with a total length of $2340$. The simulation box size was $120\times60$ in the longitudinal ($z$) and transverse ($x$) directions, respectively, with cell sizes of $\Delta z = 0.125$ and $\Delta x = 0.25$. A small time step, $\Delta t = 0.0055$, was required to resolve the electron motion and ensure sufficient data to populate the RaDiO diagnostic \cite{Vieira2021}. The total number of spatial cells was 230,400 with one particle per cell. For the radiation diagnostic RaDiO, the temporal step size was $\Delta t_R = 0.00153$, while the angular step was $0.00078 \,\text{rad}$. 

\section*{\fontsize{14}{16.8}\selectfont Acknowledgments}
\vspace{-2mm}

The authors would like to thank M. Vranic, W.B. Mori, and J. Pierce for insightful and exciting discussions.

This material is based upon work supported by the Office of Fusion Energy Sciences under Award Number DE-SC0019135 and DE-SC00215057, the Department of Energy National Nuclear Security Administration under Award Number DE-NA0003856, the University of Rochester, and the New York State Energy Research and Development Authority.


This report was prepared as an account of work sponsored by an agency of the U.S. Government. Neither the U.S. Government nor any agency thereof, nor any of their employees, makes any warranty, express or implied, or assumes any legal liability or responsibility for the accuracy, completeness, or usefulness of any information, apparatus, product, or process disclosed, or represents that its use would not infringe privately owned rights. Reference herein to any specific commercial product, process, or service by trade name, trademark, manufacturer, or otherwise does not necessarily constitute or imply its endorsement, recommendation, or favoring by the U.S. Government or any agency thereof. The views and opinions of authors expressed herein do not necessarily state or reflect those of the U.S. Government or any agency thereof.

\section*{\fontsize{14}{16.8}\selectfont Author Contributions}
\vspace{-2mm}
D.R. and J.P.P. developed the theory and wrote the paper with input from the other authors. B.M. and J.V. performed OSIRIS and RaDiO simulations. M.P. and J.V. developed RaDiO. A.D.P. and M.F. provided input into the theory and expertise regarding quantum extensions. P.F., D.H.F., B.M., T.T.S., J.V., and K.W. provided conceptual expertise for experimental considerations and the overall idea.

\bigskip
\noindent \textbf{Competing interests:} The authors declare no competing interest.

\bigskip
\noindent \textbf{Data availability:} Data and code are available on request from the authors.

\newpage
\printbibliography[title={\normalsize
References:}]

\end{document}